\documentclass[11pt,a4paper]{article}
\usepackage[cp1251]{inputenc} %Windows

\usepackage{amsmath,amsthm,amssymb,epsfig,latexsym}
\usepackage{ulem}
\usepackage[english]{babel}

 \usepackage[usenames,dvipsnames]{pstricks}
 \usepackage{epsfig}
\usepackage{pst-grad} % For gradients
\usepackage{pst-plot} % For axes

\def\Uq#1{U_q(\widehat{\mathfrak{gl}}_{#1})}
\def\Uqin#1{U_{q^{-1}}(\widehat{\mathfrak{gl}}_{#1})}
\def\rvec{|0\>}
\def\lvec{\<0|}

\def\beq#1{\begin{equation}\label{#1}}
\def\ba#1{\begin{multline}\label{#1}}
\def\eeq{\end{equation}}
\def\ea{\end{multiline}}

\def\Izer{{\sf K}}
\def\fun{{\sf f}}
\def\gun{{\sf g}}

\def\<{\langle}
\def\>{\rangle}
\def\CC{{\mathbb C}}

\def\r#1{(\ref{#1})}

\def\ot{\otimes}

\def\sk#1{\left(#1\right)}

\def\bbb{\mathbb{B}}

\def\si{\sigma}
\def\RR{{\rm R}}
\def\LL{{T}}
\def\E{{\rm E}}

\let\segg=\relax %\def\segg#1{[#1]}

\let\qsym=\beta

\def\ccc{\mathbb{C}}

\def\mu{v}

\newcommand{\Sym}{{\mathop{\rm Sym}}}

\newtheorem{prop}{Proposition}[section]

\newtheorem{exa}{Example}[section]

\newcommand{\wh}[1]{\widehat{#1}}

 \makeatletter
 \@addtoreset{equation}{section}
 \makeatother

\def\rvec{|0\>}
\def\lvec{\<0|}

\def\Izer{{\sf K}}
\def\fun{{\sf f}}
\def\gun{{\sf g}}

\def\<{\langle}
\def\>{\rangle}
\def\CC{{\mathbb C}}

\def\r#1{(\ref{#1})}

\def\ot{\otimes}

\def\sk#1{\left(#1\right)}

\def\bbb{\mathbb{B}}

\def\si{\sigma}
\def\RR{{\rm R}}
\def\LL{{T}}
\def\E{{\rm E}}

\def\ccc{\mathbb{C}}

\newcommand{\so}{{\scriptscriptstyle{\rm I}}}
\newcommand{\st}{{\scriptscriptstyle{\rm I\hspace{-1pt}I}}}

      \def\cO{{\cal O}}

\newcommand{\mb}[1]{\quad\mbox{#1}\quad}
\newcommand{\ben}{\begin{eqnarray}}
\newcommand{\een}{\end{eqnarray}}
\newcommand{\nonu}{\nonumber\\}

\newcommand{\vph}{\varphi}
\newcommand{\Ht}{\mathcal{T}}

\def\ll{\lambda}
\def\t{t}
\def\prect{\,{\prec}^t\,}

\def\gone{{\sf g}^{(l)}}
\def\gtwo{{\sf g}^{(r)}}

\newcommand{\Izerl}{\Izer^{(l)}}
\newcommand{\Izerr}{\Izer^{(r)}}

\newcommand{\II}{{\mathbb I}}
\def\vph{\varphi}

\newcommand{\finprf}{\null \hfill {\rule{5pt}{5pt}}\\[2.1ex]\indent}

\def\admis#1{[\bar #1]} % \def\admis#1{[[\bar #1]]}
\def\ms{m}
\def\ss{s}
\def\bm{\bar{\mathbf{m}}}
\def\bs{\bar{\mathbf{s}}}
\def\bfm{{\mathbf{m}}}
\def\bfs{{\mathbf{s}}}

\def\rvac{|0\rangle}

\def\hi{\hat \imath}
\def\hj{\hat \jmath}
\def\bt{\bar t}

\begin{document}

\renewcommand*{\thefootnote}{\fnsymbol{footnote}}
\begin{flushright}
LAPTH-058/13
\end{flushright}

\vspace{20pt}

\begin{center}
\begin{LARGE}
{\bf Bethe vectors of quantum integrable models\\ based on $\Uq{N}$}
\end{LARGE}

\vspace{40pt}

\begin{large}
{S.~Pakuliak${}^a$, E.~Ragoucy${}^b$, N.~A.~Slavnov${}^c$\footnote{pakuliak@theor.jinr.ru, eric.ragoucy@lapp.in2p3.fr, nslavnov@mi.ras.ru}}
\end{large}

 \vspace{12mm}

\vspace{4mm}

${}^a$ {\it Laboratory of Theoretical Physics, JINR, 141980 Dubna, Moscow reg., Russia,\\
Moscow Institute of Physics and Technology, 141700, Dolgoprudny, Moscow reg., Russia,\\
Institute of Theoretical and Experimental Physics, 117259 Moscow, Russia}

\vspace{4mm}

${}^b$ {\it Laboratoire de Physique Th\'eorique LAPTH, CNRS and Universit\'e de Savoie,\\
BP 110, 74941 Annecy-le-Vieux Cedex, France}

\vspace{4mm}

${}^c$ {\it Steklov Mathematical Institute,
Moscow, Russia}

\vspace{4mm}

\end{center}

\vspace{2mm}

\begin{abstract}
We study quantum $\Uq{N}$ integrable models solvable by the nested algebraic Bethe ansatz.
Different formulas are given for the right and left universal off-shell nested Bethe vectors.
It is shown that these formulas   can be
related by certain morphisms of the positive Borel subalgebra in $\Uq{N}$ into analogous subalgebra
in $U_{q^{-1}}(\widehat{\mathfrak{gl}}_N)$.
\end{abstract}

\vspace{2mm}

%%%%%%%%%%%%%%%%%%%%%%%%%%%%%%%%%%%%%%%%%%%%%

\renewcommand*{\thefootnote}{\arabic{footnote}}
\addtocounter{footnote}{-1}
%%%%%%%%%%%%%%%%%%%%%%%%%%%%%%%%%%%

\newpage

\section{Introduction}

The nested algebraic Bethe ansatz \cite{KulRes83,KulRes81,KulRes82} in its original formulation
allows one to get the Bethe equations as conditions that the Bethe vectors (BV) are eigenstates of the transfer
matrix. Nevertheless, even when the Bethe parameters are free and do not satisfy any restriction, the structure
of the BV (sometimes such BV are called off-shell) is rather complicated.
In the theory of solutions of the
quantum Knizhnik--Zamolodchikov equation \cite{VT} the universal off-shell  BV were given by certain trace over
auxiliary spaces of the products of the monodromy matrices and $\RR$-matrices. This presentation allows
one to investigate the structure of the nested off-shell BV. It also leads to
the explicit formulas for the nested BV when the quantum space of the integrable model becomes the tensor product
of  evaluation representations of the Yangian or of the Borel subalgebra of the quantum
affine algebra $\Uq{N}$ \cite{VTcom}.

The explicit formulas for the off-shell BV in terms of the matrix elements of the
monodromy matrix were obtained in the papers \cite{KhP-Kyoto,OPS} in the framework
of the so-called current approach. In this method the off-shell BV are identified
with projections of the product of $\Uq{N}$ currents onto intersections of the standard and current
Borel subalgebras in the quantum affine algebra $\Uq{N}$. The theory of these projections
was elaborated  in the pioneering paper \cite{ER} and then fully developed in \cite{EKhP}.

The main results of the paper  \cite{KhP-Kyoto} is a development of the
method firstly discovered in \cite{KhP-sl3}
to calculate the projections of the product of the currents in the case of
quantum affine algebra $\Uq{N}$. It was shown in \cite{OPS} that in order to obtain
different presentations of the BV associated with two different embeddings
$\Uq{N-1}$ into $\Uq{N}$, one needs to explore two different current realizations of the
quantum affine algebra $\Uq{N}$.  
It becomes clear now  that these
two different current realizations are related by certain morphisms. In the present paper we show,
in particular, how these maps allow us
to obtain the dual (or left) off-shell BV from the right ones.

It was shown in \cite{KulRes83,KulRes81,KulRes82} that the spectrum of the transfer matrix can be obtained
without {the} use of explicit formulas for BV. Such explicit formulas, however, are very important for the calculation of scalar products of BV, which, in turn, are the main tools for the analysis of correlation functions and form factors of local operators in the Bethe ansatz solvable models. To address this very complicated problem in the case of integrable models with $\Uq{N}$ symmetry, one has to get convenient formulas for the off-shell BV. Such type of presentations were obtained in \cite{BelPRS12c} for the BV in the models with $\mathfrak{gl}_3$-invariant $\RR$-matrix. There the explicit formulas for off-shell BV were given in terms of  sums over partitions of the sets of  Bethe parameters. These expressions also include the Izergin determinant \cite{Ize-det}, which is the partition function of the six-vertex model with domain wall boundary conditions \cite{Kor82}.  The properties of the Izergin determinant allow one to obtain compact formulas for the scalar products of BV in several important particular cases.  Using these formulas and the solution of the inverse scattering problem \cite{MaiTer00,KitMaiT99} we succeeded to calculate form factors of some local operators in the ${\rm SU}(3)$-invariant $XXX$ Heisenberg chain \cite{BelPRS13a}.

The main goal of this paper is to extend the results of \cite{BelPRS12c} for the models with $\Uq{N}$ symmetry.
Our starting point is the explicit formulas for the off-shell nested BV obtained in \cite{KhP-Kyoto,OPS}
in terms of the summation over permutations of the whole set of  Bethe parameters. First we observe that partial summations over permutations give the Izergin determinants. The remaining symmetrization then leads to explicit formulas
for the right and left off-shell BV in terms of  sums over partitions of the sets of  Bethe parameters.
We will obtain two different
presentations for the same right BV corresponding to the different ways of embedding $\Uq{N-1}$
into $\Uq{N}$.

Then using special properties of the trigonometric $\RR$-matrix, we will define the special
morphism of $\Uq{N}$ to $\Uqin{N}$ and prove that different presentations for the right off-shell BV
can be related by this morphism. Further we will define one more antimorphism of $\Uq{N}$ to $\Uqin{N}$
which allows us to obtain the formulas in terms of  sums over partitions for the left (or dual)
off-shell BV.

\subsection{Notations}

In this paper we consider the quantum integrable models defined by  the $N\times N$
monodromy matrix $\LL_{ij}(z)$ satisfying the commutation relation
\begin{equation}\label{YB}
\RR(u,v;q)\cdot (\LL(u)\ot \mathbf{1})\cdot (\mathbf{1}\ot \LL(v))=
(\mathbf{1}\ot \LL(v))\cdot (\LL(u)\ot \mathbf{1})\cdot \RR(u,v;q)\,,
\end{equation}
where  $\RR(u,v;q)\in{\textrm{End}}(\CC^N\ot\CC^N)\ot \CC[[{v}/{u}]]$,
is  a trigonometric $\RR$-matrix associated with the vector
representation of $\Uq{N}$.
Here  $q$ is  a complex parameter neither equal to zero
  nor root of unity.
 The algebra \r{YB} describes also the commutation relation in the
standard Borel subalgebra of the quantum affine algebra $\Uq{N}$.
In what follows we will describe the morphisms of this subalgebra of $\Uq{N}$ into
the analogous subalgebra of $\Uqin{N}$ and this is a reason why we are writing
the explicit dependence of the $\RR$-matrix on the parameter $q$.

More explicitly this $\RR$-matrix can be written in the form
\begin{equation}\label{UqglN-R-two}
\begin{split}
\RR(u,v;q)\ =\ \fun_q(u,v)&\ \sum_{1\leq i\leq N}\E_{ii}\ot \E_{ii}\ +\
\sum_{1\leq i<j\leq N}(\E_{ii}\ot \E_{jj}+\E_{jj}\ot \E_{ii})
\\
+\ &\sum_{1\leq i<j\leq N}
\big(\gone_q(u,v) \E_{ij}\ot \E_{ji}+ \gtwo_q(u,v)\E_{ji}\ot \E_{ij}\big)\,,
\end{split}
\end{equation}
where
$(\E_{ij})_{lk}=\delta_{il}\delta_{jk}$, $i,j,l,k=1,...,N$ are $N\times N$ matrices
with unit in the intersection of $i$th row and $j$th column and zero  elsewhere,
and the coefficient functions are defined as follows\footnote{When there is no ambiguity, we will omit the subscript $q$ in the rational functions \r{fqq} to simplify the formulas.}
\begin{equation}\label{fqq}
\fun_q(u,v)=\frac{qu-q^{-1}v}{u-v},\quad \gun_q(u,v)=\frac{(q-q^{-1})}{u-v},\quad
\end{equation}
and
\begin{equation}\label{glr}
\gone_q(u,v)=u\gun_q(u,v)\,,\quad \gtwo_q(u,v)=v\gun_q(u,v)\,.
\end{equation}

A concrete quantum integrable model is defined usually by a certain representation space $V$ (or quantum space
of the integrable models) where
the {entries} of the monodromy matrix $\LL_{i,j}(u)$ {act}.
The space of states  for this model is identified with a representation space
of the Borel subalgebra of $\Uq{N}$ generated by the vector $\rvec$ satisfying the
following conditions
\begin{equation}\label{rvec}
\LL_{j,i}(z)\rvec=0,\quad j>i,\quad \LL_{i,i}(z)\rvec=\ll_i(z)\rvec,\quad i=1,\ldots,N\,.
\end{equation}
The functions $\ll_i(z)$ {characterize} the concrete integrable model and the vector $\rvec$
is a common eigenvector of the diagonal {entries} of the monodromy matrix. Although the diagonal
matrix elements do not commute, such a vector exists because the commutators
$[\LL_{i,i}(z),\LL_{j,j}(z')]$, $\forall i,j$,
annihilate the vacuum vector $\rvec$, due to the commutation relations and   \r{rvec}.

The off-shell BV are constructed as special polynomials of the monodromy
matrix elements $\LL_{i,j}(z)$, $i\leq j$, depending on  sets of parameters (the Bethe parameters)
and acting on $\rvec$. These parameters are supposed to be generic complex numbers. If they satisfy the system of the
nested Bethe ansatz equations, then the corresponding BV becomes an eigenvector of the transfer matrix\footnote{%
Sometimes such vectors are called on-shell BV.}.

We will also consider the dual (or left) off-shell BV, which belong to the dual space $V^*$. They are
generated by the matrix elements  $\LL_{i,j}(z)$  acting on a vector $\lvec$, which satisfies the conditions
\begin{equation}\label{lvec}
\lvec \LL_{i,j}(z)=0,\quad j>i,\quad \lvec\LL_{i,i}(z)=\ll_i(z)\lvec,\quad i=1,\ldots,N\,.
\end{equation}

\section{Formulas for BV from the current approach}

The goal of this section is to remind the formulas for the right off-shell BV formulated
by the Theorem 2 of the paper \cite{OPS} in the form of sums over all permutations of the
Bethe parameters of the same sort.

Let  $\segg{\bar n}=\{n_1,n_2,\ldots,n_{N-1}\}$ be a collection of positive integers and
let
$\bar{t}_{\segg{\bar{n}}} $ be a set of  variables
\begin{equation}\label{set111}
\bar{t}_{\segg{\bar{n}}} = \left\{
t^{1}_{1},\ldots,t^{1}_{n_{1}}; t^{2}_{1},\ldots,
t^{2}_{n_{2}}; \ldots\ldots; t^{N-2}_{1},\ldots, t^{N-2}_{n_{N-2}};
t^{N-1}_{1},\ldots,t^{N-1}_{n_{N-1}}\right\}\,.
\end{equation}
Here superscripts indicate the type of  Bethe parameter  and correspond
to the simple roots of the algebra $\mathfrak{gl}_N$. There are $N-1$ different sorts of Bethe parameters
in the generic BV for $\Uq{N}$-integrable model. The subscript counts the number of the
Bethe parameters of the same type. For a generic BV, we denote by $n_i$ the total number
of  type $i$ Bethe parameters.

Let us consider
a direct product of the symmetric groups:
$S_{\bar n} =S_{n_1}\times \cdots
\times S_{n_{N-1}}$.
For any  function $G(\bar t_{\segg{\bar n}})$
we denote by
\begin{equation}\label{symmet}
\Sym_{\ \bar t_{\segg{\bar n}}} \ G(\bar t_{\segg{\bar n}})= \sum_{\si\in
S_{\bar n}}\ G(^\si \bar t_{\segg{\bar n}})\,,\quad  \sigma=\{\sigma^1,\sigma^2,\ldots,\sigma^{N-1}\}
\end{equation}
a symmetrization over groups of variables of same type
$k$, $\{t^k_1,\ldots,t^k_{n_k}\}$, where
\begin{equation}\label{sigmat}
^\si \bar t_{\segg{\bar n}} =\{t^{1}_{\si^{1}(1)},\ldots,
t^{1}_{\si^{1}(n_1)};
\ldots;t^{N-1}_{\si^{N-1}(1)},\ldots,t^{N-1}_{\si^{N-1}(n_{N-1})}\}.
\end{equation}
Let
\begin{equation*}%\label{omega-f}
\qsym(\bar t_{\segg{\bar n}})=\prod_{k=1}^{N-1}\prod_{1\leq \ell<\ell'\leq n_k}
\fun(t^k_{\ell'},t^k_\ell)
\end{equation*}
be a function of the formal variables $t^k_\ell$, $\ell=1,..,n_k$, $k=1,...,N-1$.

In order to describe formulas for the off-shell BV we need the following
combinatorial data.
Let $\admis{\ms}=\{\ms_j^i\}$ and
$\admis{\ss}=\{\ss_i^j\}$
for $1\leq i\leq j\leq N-1$
be two collections of the nonnegative integers.
We say that collections  $\admis{\ms}$ and $\admis{\ss}$ are $\bar{n}$ upper or lower (resp.) permissible,
if they satisfy the following conditions
\begin{equation}\label{admis-m}
\ms^i_i\geq \ms^i_{i+1}\geq\cdots\geq \ms^i_{N-1}\geq \ms^i_N=0\,,
\quad n_i=\sum_{j=1}^{i}\ms_i^j\,,\quad i=1,\ldots, N-1\,,
\end{equation}
and
\begin{equation}\label{admis-h}
0=\ss^i_0\leq \ss^i_1\leq\cdots\leq \ss^{i-1}_{i-1}\leq \ss^i_i\,,\quad n_i=\sum_{j=i}^{N-1}\ss_i^j\,,\quad
i=1,\ldots, N-1\,,
\end{equation}
respectively.
We also follow the convention $\ms^j_N=\ss_0^j=0$ for $j=1,...,N-1$.

The collections of upper or lower permissible integers $\admis{\ms}$ and
$\admis{\ss}$ can be visualized as  upper or lower triangular matrices
\begin{equation}\label{mad-mat}
\admis{\ms}=
\left(
\begin{array}{ccccc}
\ms_{1}^{1}& \ms_{2}^{1}&\ldots&\ms_{N-2}^{1}&\ms_{N-1}^{1}\\[2mm]
0&\ms_{2}^{2}&\ldots&\ms_{N-2}^{2}&\ms_{N-1}^{2}\\[2mm]
&&\ddots&\vdots&\vdots\\[2mm]
&0&&\ms^{N-2}_{N-2}&\ms^{N-2}_{N-1}\\[2mm]
&&&&\ms_{N-1}^{N-1}
\end{array}
\right)
\
\begin{array}{c}
0=\ms_N^{1}\!\!\leq \ms^{1}_{N-1}\leq \ldots\,\leq \ms_{1}^{1}\,,  \\[2mm]
0=\ms_N^{2}\!\!\leq \ms_{N-1}^{2}\leq \ldots \!\leq \ms_{2}^{2}\,, \\[2mm]
\vdots\\[2mm]
0=\ms^{N-2}_N\leq \ms^{N-2}_{N-1}\leq \ms^{N-2}_{N-2}\,,\\[2mm]
0=\ms^{N-1}_N\leq \ms_{N-1}^{N-1}\,,
\end{array}
\end{equation}
and
\begin{equation}\label{ad-mat}
\admis{s}=
\left(
\begin{array}{ccccc}
s_{1}^{1}&&&&\\[2mm]
s^{2}_1&s^{2}_{2}&&0&\\[2mm]
\vdots&\vdots&\ddots&&\\[2mm]
 s_{1}^{N-2}&s_{2}^{N-2}&\ldots&s_{N-2}^{N-2}&\\[2mm]
s_{1}^{N-1}& s_{2}^{N-1}&\ldots&s_{N-2}^{N-1}&s_{N-1}^{N-1}
\end{array}
\right)\
\begin{array}{c}
0=s^1_0\leq s_{1}^{1}\,, \\[2mm]
0=s^2_0\leq s^{2}_{1}\leq s^{2}_2\,,\\[2mm]
\vdots\\[2mm]
0=s_0^{N-2}\!\!\leq s_{1}^{N-2}\leq \ldots \!\leq s_{N-2}^{N-2}\,,   \\[2mm]
0=s_0^{N-1}\!\!\leq s_{1}^{N-1}\leq \ldots\,\leq s_{N-1}^{N-1}\,.
\end{array}
\end{equation}
In words, $\admis{\ms}$ is an upper triangular matrix with integer entries, ordered on each lines, and such that the sum of its $i^{{th}}$ column gives back $n_i$.
In what follows, $\admis{\ms}$ (resp. $\admis{\ss}$) will denote upper (resp. lower) permissible collections of integers.

Let $\bar\ms^j$ and $\bar\ss^j$, $j=1,\ldots,N-1$ be the $j$-th rows of the permissible matrices
  $\admis{\ms}$ and $\admis{\ss}$.
Define a collection of vectors
\begin{equation}\label{mi-def}
\bm^j=\bar\ms^{1}+\bar\ms^{2}+\cdots+\bar\ms^{j-1}+\bar\ms^{j}\,,\quad
 j=1,\ldots,N-1\,,
\end{equation}
\begin{equation}\label{si-def}
\bs^j=\bar\ss^{j}+\bar\ss^{j+1}+\cdots+\bar\ss^{N-2}+\bar\ss^{N-1}\,,\quad
 j=1,\ldots,N-1\,,
\end{equation}
with non-negative integer components.
Set $\bm^0=\bar 0$ and $\bs^N=\bar 0$.
Denote by $\bfm^j_a$ and $\bfs^j_a$
the components of the vectors
$\bm^j$ and $\bs^j$:
\begin{equation*}
\bm^j=\{n_1,n_2,\ldots,n_j,\ms^{1}_{j+1}+\cdots+\ms^{j}_{j+1},\ldots,
\ms^{1}_{N-1}+\cdots+\ms^{j}_{N-1}\}\,,
\end{equation*}
\begin{equation*}
\bs^j=\{\ss^{j}_{1}+\cdots+\ss^{N-1}_{1},\ldots,\ss^{j}_{j-1}+\cdots+\ss^{N-1}_{j-1},
n_{j},\ldots,n_{N-2},n_{N-1}\}\,.
\end{equation*}
According to the
conditions \r{admis-h} and \r{admis-m}, $\bm^{N-1}=\bs^1=\bar n$.

We introduce the following combinations of the
$\Uq{N}$ monodromy matrix elements, that we call pre-BV:
\begin{equation}\label{mWt3}
\begin{split}
\mathcal{B}^{\bar n}(\bar{t}_{\segg{\bar{n}}})&=
\sum_{\admis{\ms}} \Sym_{\ \bar t_{\segg{\bar n}}}
\Bigg(\qsym(\bar t_{\segg{\bar n}})
 \prod_{1\leq j\leq i< N} \bigg(  [(\ms^j_i-\ms^j_{i+1})!]^{-1}
\!\!\!\!\!\!\!\!\!\!\!\!\!\!\! \prod_{n_j-m^j_i<\ell'<\ell\leq n_j-m^j_{i+1}} \fun(t^j_\ell,t^j_{\ell'})^{-1}\bigg)\\
 &\times \prod_{i=2}^{N-1}\bigg(\prod_{j=1}^{i-1}\bigg(\prod_{\ell=0}^{\ms^j_i-1}
\gone(t^{i}_{\bfm^{j}_{i}-\ell}, t^{i-1}_{\bfm^{j}_{i-1}-\ell})
\prod_{\ell'=\bfm^j_{i-1}-\ell+1}^{n_{i-1}}
\fun(t^{i}_{\bfm^{j}_{i}-\ell},t^{i-1}_{\ell'})\bigg)\bigg)\\
 &\times
\prod_{1\leq j\leq N-1}^{\longrightarrow}\Bigg(
\prod_{N-1\geq i\geq j}^{\longleftarrow}\Bigg(
\prod_{\ell= n_j-\ms^j_i+1}^{n_j-\ms^j_{i+1}}
\LL_{j,i+1}(t^j_\ell) \Bigg)\Bigg)
\prod_{j=1}^{N-1}\prod_{\ell=1}^{n_j-\ms^j_j}\LL_{j,j}(t^j_\ell) \Bigg)\,,
\end{split}
\end{equation}
and
\begin{equation}\label{Wt3}
\begin{split}
\wh{\mathcal{B}}^{\bar n}(\bar{t}_{\segg{\bar{n}}})&=
\sum_{\admis{\ss}}\Sym_{\ \bar t_{\segg{\bar n}}}
\Bigg(\qsym(\bar t_{\segg{\bar n}})
\prod_{1\leq i\leq j< N} \bigg(  [(\ss^j_i-\ss^j_{i-1})!]^{-1}
\!\!\!\!\! \prod_{s^j_{i-1}<\ell'<\ell\leq s^j_{i}} \fun(t^j_\ell,t^j_{\ell'})^{-1}\bigg)
\\
&\times \prod_{j=2}^{N-1}\bigg(\prod_{i=1}^{j-1}\bigg(\prod_{\ell=1}^{s^j_i}
\gtwo(t^{i+1}_{n_{i+1}-\bfs^{j}_{i+1}+\ell}, t^{i}_{n_i-\bfs^{j}_{i}+\ell})
\prod_{\ell'=1}^{n_{i+1}-\bfs^j_{i+1}+\ell-1}
\fun(t^{i+1}_{\ell'}, t^{i}_{n_i-\bfs^{j}_{i}+\ell}) \bigg)\bigg)\\
 &\times
\prod_{N-1\geq j\geq 1}^{\longleftarrow}\Bigg(
\prod_{1\leq i\leq j}^{\longrightarrow}\Bigg(
\prod_{\ell= \ss^j_{i-1}+1}^{\ss^j_{i}}
\LL_{i,j+1}(t^j_\ell) \Bigg)\Bigg)
\prod_{j=1}^{N-1}\prod_{\ell=\ss^j_j}^{n_j}\LL_{j+1,j+1}(t^j_\ell) \Bigg)\,,
\end{split}
\end{equation}
where the ordered products of the non-commuting entries $A_i$ are defined as follows
\begin{equation*}
\prod_i^{\longleftarrow}\ A_i = A_nA_{n-1}\cdots A_2A_1\quad\mbox{and}\quad
\prod_i^{\longrightarrow}\ A_i=A_1A_2\cdots A_{n-1}A_n.
\end{equation*}
Theorem 2 in \cite{OPS} states that the two pre-BVs
produce two different presentations of the same off-shell BV:
\begin{equation}\label{BV1}
\bbb^{\bar n}(\bar{t}_{\segg{\bar{n}}})=\mathcal{B}^{\bar n}(\bar{t}_{\segg{\bar{n}}})\rvac
=\wh{\mathcal{B}}^{\bar n}(\bar{t}_{\segg{\bar{n}}})\rvac\,.
\end{equation}
The ordering in the products over $\ell$ in the formulas \r{mWt3}
and \r{Wt3} is not important, because of the commutativity
of the entries of $\LL$-operators with equal matrix indices.
Moreover, the order in the products of the noncommuting diagonal monodromy matrix elements
in \r{mWt3} and \r{Wt3} is also non-important, since these combinations will eventually
act onto the highest weight vector $\rvec$ and  produce products of scalar functions, due to \r{rvec}.

The formulas \r{mWt3} and \r{Wt3} contain summations over permutations of the symmetric
functions and these summations can be calculated. This will be done in the following
section and formulas for the pre-BV will be presented as sums over different partitions of the
set of  Bethe parameters  \r{set111}.

\section{Off-shell BV as sums over partitions}

To save space and simplify formulas,  we will use following convention for the products of the
{commuting entries of the monodromy matrix $T_{ij}$, vacuum eigenvalues $\lambda_i$ and functions $\fun(u,v)$.}
Namely, whenever {such an operator or a function} depends on a set of variables, say $\bar t$, this means that we
deal with the product of {these commuting operators or functions}  with respect to the corresponding set:{
 \begin{equation}\label{SH-prod}
 \LL_{i,j}(\bar t)=\prod_{t_k\in\bar t} \LL_{i,j}(t_k)\,,\qquad
 \lambda_{i}(\bar t)=\prod_{t_k\in\bar t} \lambda_{i}(t_k)\,,\qquad
  \fun(\bar t,\bar t')=\prod_{t_j\in\bar t}\ \ \ \prod_{t_k\in\bar t'} \fun(t_j,t_k)\,.
 \end{equation}
}

In various formulas the Izergin determinant $\Izer_n(\bar x|\bar y)$ appears
\cite{Ize-det}. It is defined
for two sets $\bar x$ and $\bar y$ of the same cardinality $\#\bar x=\#\bar y=n$:
\begin{equation}\label{Izer}
\Izer_n(\bar x|\bar y)=\frac{\prod_{1\leq i,j\leq k}(qx_i-q^{-1}y_j)}
{\prod_{1\leq i<j\leq k}(x_i-x_j)(y_j-y_i)}
\cdot\det \left[\frac{q-q^{-1}}{(x_i-y_j)(qx_i-q^{-1}y_j)}\right]\,.
\end{equation}
Below we also use two modifications of the Izergin determinant
\begin{equation}\label{Mod-Izer}
\Izerl_n(\bar x|\bar y)= \prod_{i=1}^nx_i\cdot\Izer_n(\bar x|\bar y)\,, \qquad
\Izerr_n(\bar x|\bar y)= \prod_{i=1}^ny_i\cdot\Izer_n(\bar x|\bar y)\,,
\end{equation}
which we call left and right Izergin determinants respectively.

\subsection{Combinatorial description of the partitions}

The main goal of this section is to transform the equations \r{mWt3} and \r{Wt3} for pre-BV into new representations
involving  sums with respect to partitions of the Bethe parameters. Let us first  describe the general strategy
of these transforms. 

Consider for definiteness \r{mWt3}. This representation contains  sums of different types. One  sum runs over all possible
permissible sets $\admis{m}$. Each individual term in this sum  includes  summations over permutations of the Bethe parameters of the same type. For example, the extreme permissible set in \r{mWt3} such that $m_i^j=\delta_i^j n_i$ corresponds
to the term
\begin{equation}\label{ex-term}
\frac{1}{n_1!n_2!\cdots n_{N-1}!}\
\Sym_{\ \bar t_{\segg{\bar n}}}\sk{\LL_{1,2}(\bar t^1)\LL_{2,3}(\bar t^2)\cdots  \LL_{N-1,N}(\bar t^{N-1})}\,,
\end{equation}
and the summation over permutations of the Bethe parameters can be easily performed due to the commutativity
of the entries of the monodromy matrix  with the same matrix indices. It is clear that this summation
removes the combinatorial factors in \r{ex-term}.

Other terms in the sum over permissible sets in \r{mWt3}  can be treated
similarly. For any permissible set $\admis{m}$  one can consider the sum over
permutations of the Bethe parameters of the same type as the sum over special partitions of these parameters into subsets  and further permutations within every subset. The partitions of the Bethe parameters into subsets are determined by the permissible set $\admis{m}$. 
 They correspond to Bethe parameters that enter
the products in \r{mWt3} through the same monodromy matrix entry.
Due to the commutativity of such matrix elements the sums over permutations inside these subsets
can be calculated via the  identities \r{ident1} and \r{ident2}. They also
remove the combinatorial factors. Details of this calculation will be presented below
in the proof of the Proposition~\ref{p31}.

The presentation \r{Wt3} can be transformed in the similar manner.
As a result we will be left with summations over all possible partitions of the sets of Bethe parameters
dictated by the permissible set $\admis{m}$ and $\admis{s}$. It is convenient to parameterize
the subsets of every set $\bar t^k$ by two positive integers $i$ and $j$ satisfying the conditions
\begin{equation}\label{pt11}
1\leq i\leq k\leq j\leq N-1\,,\quad
\bar\t^k= \mathop{\bigcup}\limits_{i=1}^k\,\mathop{\bigcup}\limits_{j=k}^{N-1}\ \bar\t^k_{i,j}\,,
\end{equation}
and
\begin{equation}\label{c11}
  \#\bar\t^k_{i,j}=m^i_j-m^i_{j+1}\mb{for} i=1,..., k\mb{and} j=k,..., N-1\,,\quad \forall k
\end{equation}
for the upper permissible matrix $\admis{m}$.  For the lower permissible matrix $\admis{s}$, we introduce analogous partitions  of the same sets \r{pt11}
with similar conditions:
\begin{equation}\label{c12}
 \#\bar\t^k_{i,j}=s^j_i-s^j_{i-1}\mb{for} i=1,..., k\mb{and} j=k,..., N-1\,,\quad \forall k\,.
\end{equation}

Note that for any given {pair} $(i,j)$, we have $\#\bar\t^k_{i,j}=\#\bar\t^{k'}_{i,j}$, $\forall k,k'$. Remark also that,
because of the conditions \eqref{admis-m} and \eqref{admis-h}, we have indeed $\#\bar\t^k=\sum_{i=1}^k\sum_{j=k}^{N-1} \#\bar\t^k_{i,j}$.

We introduce ordering rules  '$\prec$' and '$\prect$' of these pairs according to the following conventions
\begin{equation}\label{rela1}
i,j\prec i',j'\quad  \mbox{if}\quad i<i',\ \ \forall j,j'\quad\mbox{or if}\quad i=i', j<j'\,,
\end{equation}
and
\begin{equation}\label{rela2}
i,j\prect  i',j'\quad \mbox{if}\quad j<j',\ \ \forall i,i'\quad\mbox{or if}\quad j=j', i<i'\,.
\end{equation}

\begin{exa}
To illustrate how combinatorial data encoded into permissible matrices \r{mad-mat} and \r{ad-mat}
is transferred into description of partitions \r{pt11}
we write explicitly the partitions associated with this data in case of the $\Uq{5}$ off-shell BV.

In that case, we have 4 types of Bethe parameters $\bar t^1,...,\bar t^4$, and the permissible collections lead to $4\times 4$ triangular matrices of types \r{mad-mat} and \r{ad-mat}.
The  rules  \r{pt11} show that the sets of Bethe parameters $\bar t^1$ and $\bar t^4$ (resp. $\bar t^2$ and $\bar t^3$) are divided into 4 (resp. 6) subsets. We get the subsets
\begin{equation*}
\begin{array}{cccccccccccccccccccc}
\bt^1=&\bt^1_{1,1}&\cup&\bt^1_{1,2}&\cup&\bt^1_{1,3}&\cup&\bt^1_{1,4}&&&&&&&&&&&&\\[3mm]
\bt^2=&&&\bt^2_{1,2}&\cup&\bt^2_{1,3}&\cup&\bt^2_{1,4}&\cup&\bt^2_{2,2}&\cup&\bt^2_{2,3}&\cup&\bt^2_{2,4}     &&&&&&\\[3mm]
\bt^3=&&&&&\bt^3_{1,3}&\cup&\bt^3_{1,4}&\cup&&\cup&\bt^3_{2,3}&\cup&\bt^3_{2,4}&\cup&\bt^3_{3,3}&\cup&\bt^3_{3,4}&&\\[3mm]
\bt^4=&&&&&&&\bt^4_{1,4}&\cup&&&&\cup&\bt^4_{2,4}&\cup&&\cup&\bt^4_{3,4}&\cup&\bt^4_{4,4}
\end{array}
\end{equation*}
for \r{mad-mat}   and the subsets
\begin{equation*}
\begin{array}{cccccccccccccccccccc}
\bt^4=&\bt^4_{4,4}&\cup&\bt^4_{3,4}&\cup&\bt^4_{2,4}&\cup&\bt^4_{1,4}&&&&&&&&&&&&\\[3mm]
\bt^3=&&&\bt^3_{3,4}&\cup&\bt^3_{2,4}&\cup&\bt^3_{1,4}&\cup&\bt^3_{3,3}&\cup&\bt^3_{2,3}&\cup&\bt^3_{2,4}     &&&&&&\\[3mm]
\bt^2=&&&&&\bt^2_{2,4}&\cup&\bt^2_{1,4}&\cup&&\cup&\bt^3_{2,3}&\cup&\bt^3_{1,3}&\cup&\bt^3_{2,2}&\cup&\bt^2_{1,2}&&\\[3mm]
\bt^1=&&&&&&&\bt^1_{1,4}&\cup&&&&\cup&\bt^1_{1,3}&\cup&&\cup&\bt^1_{1,2}&\cup&\bt^1_{1,1}
\end{array}
\end{equation*}
for \r{ad-mat}. The subsets in {the same} column have the same cardinality, {that is} defined by the formulas \r{c11} and \r{c12}.
\end{exa}

Now we can formulate the following
\begin{prop}\label{p31}
The formulas \r{mWt3} and \r{Wt3} for the universal pre-BV can be written as sums over
partitions of the Bethe parameters \r{pt11}, \r{c11} and \r{c12} as follows
\begin{equation}\label{bvp1}
\begin{split}
\mathcal{B}^{\bar n}(\bar{t}_{\segg{\bar{n}}})&=\sum_{{\rm part}}
\prod_{k=1}^{N-1}\prod_{i,j\prec i',j'} \fun(\bar\t^k_{i',j'},\bar\t^k_{i,j})
\prod_{k=2}^{N-1}\Bigg(\prod_{i,j\prec i',j'} \fun(\bar \t^k_{i,j},\bar \t^{k-1}_{i',j'})
\prod_{i<j} \Izerl(\bar \t^k_{i,j}|\bar \t^{k-1}_{i,j})\Bigg)\\
&\times \prod_{1\leq k\leq N-1}^{\longrightarrow} \Bigg(\prod_{N\geq j> k}^{\longleftarrow}
\LL_{k,j}(\bar\t^k_{k,j-1})\Bigg)\prod_{k=2}^{N-1}\prod_{i,j\prec k,k}\LL_{k,k}(\bar\t^k_{i,j})\,,
\end{split}
\end{equation}
and
\begin{equation}\label{bvp2}
\begin{split}
\wh{\mathcal{B}}^{\bar n}(\bar{t}_{\segg{\bar{n}}})&=\sum_{{\rm part}}
\prod_{k=1}^{N-1}\prod_{i,j\prect i',j'} \fun(\bar\t^k_{i',j'},\bar\t^k_{i,j})
\prod_{k=2}^{N-1}\Bigg(\prod_{i,j\prect i',j'} \fun(\bar \t^k_{i,j},\bar \t^{k-1}_{i',j'})
\prod_{i<j} \Izerr(\bar \t^k_{i,j}|\bar \t^{k-1}_{i,j})\Bigg)\\
&\times \prod_{N-1\geq k\geq 1}^{\longleftarrow}\Bigg(\prod_{1\leq j\leq k}^{\longrightarrow}
\LL_{j,k+1}(\bar\t^k_{j,k})\Bigg)
\prod_{k=1}^{N-2}\prod_{k,k\prect i,j}\LL_{k+1,k+1}(\bar\t^k_{i,j})\,.
\end{split}
\end{equation}
\end{prop}
\proof
The basic idea  is to replace the summations over permutations of the Bethe parameters and
permissible matrices $\admis{m}$, $\admis{s}$ in \r{mWt3} and \r{Wt3} by the summations
over partitions of the sets of the Bethe parameters.  Once it is done, the sum over permutations within any fixed  subset can be calculated using certain identities of the
rational functions \r{ident1} and \r{ident2}. To formulate these identities we introduce the following notations.
For any sets $\bar x$ and $\bar y$ with equal cardinalities
$\#{\bar x}=\#{\bar y}=n$, we {introduce} rational functions
\begin{equation}\label{not1}
{\sf B}(\bar y)=\prod_{\ell<\ell'}\fun(y_{\ell'},y_\ell),\quad
{\sf G}(\bar y|\bar x)=\prod_{\ell=1}^n\gun(y_\ell,x_\ell),\quad
{\sf F}(\bar y|\bar x)=\prod_{\ell<\ell'}\fun(y_{\ell},x_{\ell'}).
\end{equation}
Then the following identities of the rational functions are valid:
\begin{equation}\label{ident1}
\Sym_{\bar y}\big({\sf B}(\bar y){\sf G}(\bar y|\bar x){\sf F}(\bar y|\bar x)\big)=\Izer_n(\bar y|\bar x)\,,
\end{equation}
\begin{equation}\label{ident2}
\Sym_{\bar x}\big({\sf B}(\bar x){\sf G}(\bar y|\bar x){\sf F}(\bar y|\bar x)\big)=\Izer_n(\bar y|\bar x)\,.
\end{equation}
The proof of these identities in the case of  quantum integrable models associated with
rational $\RR$-matrix is given in Appendix A of the paper \cite{BelPRS12a}. The proof in the
trigonometric case is {completely} analogous.

According to the  partitions defined by the permissible matrices $\admis{m}$ and $\admis{s}$
the rational function $\qsym(\bar t_{\segg{\bar n}})$ can be rewritten as
\begin{equation}\label{part1}
\qsym(\bar t_{\segg{\bar n}})=\prod_{k=1}^{N-1}\prod_{1\leq \ell<\ell'\leq  n_k}\fun(t^k_{\ell'},t^k_\ell)=
\prod_{k=1}^{N-1}{\sf B}({\bar t}^k)=
\prod_{i\leq k\leq j}{\sf B}({\bar t}^k_{i,j}),
\end{equation}
where ${\bar t}^k_{i,j}$ are defined either by  \r{c11} or by  \r{c12}.

We show the equivalence of \r{mWt3} and \r{bvp1} first,  using partitions \r{c11}. We will decompose the proof in the
series of steps.
\begin{itemize}
\item[{\bf 1.}] The product of rational functions $\prod_{i\leq j}^{N-1} {\sf B}(\bar t^i_{i,j})$
entering the function $\qsym(\bar t_{\segg{\bar n}})$ will be canceled out by the product of the
rational functions in the first row of \r{mWt3}. This allows to cancel some of the factorial
factors. At this step the expression in \r{mWt3} is symmetric with respect to permutations
in the set $\bar t^{N-1}_{N-1,N-1}$ and the factorial factor $\big(m^{N-1}_{N-1}!\big)^{-1}$ disappear.
\item[{\bf 2.}]  At this step, for $k=1,\ldots,N-2$, we can select the following product of  rational functions
in the r.h.s. of \r{mWt3}
\begin{equation*}
 {\sf B}(\bar t^{N-1}_{k,N-1}){\sf G}(\bar t^{N-1}_{k,N-1}|\bar t^{N-2}_{k,N-1})
 {\sf F}(\bar t^{N-1}_{k,N-1}|\bar t^{N-2}_{k,N-1}).
\end{equation*}
Since all the other factors in the r.h.s. of \r{mWt3} are symmetric with respect to the permutations
in the set $\bar t^{N-1}_{k,N-1}$, we can perform the symmetrization over these sets and apply the identity
\r{ident1} to obtain the product over $k$ of the Izergin determinants
$\Izer(\bar t^{N-1}_{k,N-1}|\bar t^{N-2}_{k,N-1})$ which are symmetric  with respect
to permutations in the sets $\bar t^{N-2}_{k,N-1}$. Moreover the factorial factors
$\Big((m^{N-2}_{N-2}-m^{N-2}_{N-1})!\,m^{N-2}_{N-1}!\Big)^{-1}$
will disappear since the expression under summation
is symmetric with respect to permutations in the sets $\bar t^{N-2}_{N-2,N-2}$ and $\bar t^{N-2}_{N-2,N-1}$.
\item[{\bf 3.}] Next we perform an analogous procedure to obtain the product of  the Izergin
determinants over $k=1,\ldots,N-3$
\begin{equation*}
\Izer(\bar t^{N-2}_{k,N-2}|\bar t^{N-3}_{k,N-2})\Izer(\bar t^{N-2}_{k,N-1}|\bar t^{N-3}_{k,N-1})\,,
\end{equation*}
which is symmetric with respect to permutations in the sets $\bar t^{N-3}_{k,N-1}$ and
$\bar t^{N-3}_{k,N-2}$. Now the combinatorial factors
$\Big((m^{N-3}_{N-3}-m^{N-3}_{N-2})!\,(m^{N-3}_{N-2}-m^{N-3}_{N-1})!\,m^{N-3}_{N-1}!\Big)^{-1}$
will disappear due to the symmetry of the summand with respect to permutations in the sets
$\bar t^{N-3}_{N-3,N-3}$, $\bar t^{N-3}_{N-3,N-2}$ and $\bar t^{N-3}_{N-3,N-1}$.
\item[{\bf 4.}]  We iterate these symmetrizations over the sets up to the final step,
that corresponds to the symmetrizations
over the sets $\bar t^2_{1,k}$, $k=2,\ldots,N-1$. We obtain in this way the product of
Izergin determinants
$\prod_{k=2}^{N-1}\Izer (\bar t^2_{1,k}|\bar t^1_{1,k})$
and the last combinatorial factor $\prod_{k=1}^{N-1}((m^1_k-m^1_{k+1})!)^{-1}$ disappears,
due to symmetry arguments.
\end{itemize}

Besides the product of  Izergin determinants described above, we are also left with the product
of rational functions, which can we written as in the first line of \r{bvp1} using the
ordering of the sets introduced by \r{rela1}.

The proof of equivalence between \r{Wt3} and \r{bvp2} is analogous. The only difference is that
we have to start the whole procedure  with the partitions \r{c12}, and begin with the symmetrizations over sets $\bar t^1_{1,1}$, $\bar t^1_{1,2}$ and so on. Then,
we use the identity \r{ident2} to produce the product of the Izergin determinants.
\finprf

\section{Morphisms}

The aim of this section is to describe certain morphisms which relate
the algebras $\Uq{N}$ and $\Uqin{N}$ \cite{CP94}.
It will be shown that one of these  morphisms relate pre-BV \r{bvp1} and \r{bvp2}.
Using second morphism one can obtain dual or left pre-BV in the form of sums over
partitions of the Bethe parameters from the formulas for the right pre-BV \r{bvp1} and \r{bvp2}.

\subsection{Properties of the $R$-matrix}

Note that the functions \r{fqq} introduced above,
and the modified Izergin determinants \r{Mod-Izer} obey the following relations
\begin{equation}\label{prop-fct}
\begin{split}
\gone_{q^{-1}}(v,u)=\gtwo_q(u,v),\qquad&\gone_{q^{-1}}(u^{-1},v^{-1})=\gtwo_q(u,v),
\\
\fun_{q^{-1}}(v,u)=\fun_q(u,v),\qquad&\fun_{q^{-1}}(u^{-1},v^{-1})=\fun_q(u,v),
\\
\Izer^{(l)}_{q^{-1}}(\bar v|\bar u)=\Izer^{(r)}_q(\bar u|\bar v),
\qquad&\Izer^{(l)}_{q^{-1}}(\bar u^{-1}|\bar v^{-1})=\Izer^{(r)}_q(\bar u|\bar v).
\end{split}
\end{equation}

We {also} have the following properties, that {can be} proved by direct calculation:
\beq{eq:symR}
\begin{split}
\RR_{12}(u,v)\,\RR_{21}(v,u)=\fun_q(u,v\,)\fun_q(v,u)\, \II\otimes\II,\qquad & \RR_{12}(u,v)^{t_1t_2}\ =\ \RR_{21}(u,v),
\\
U_1\,U_2\,\RR_{12}(u,v)\,U_1^{-1}\,U_2^{-1}\ =\ \RR_{21}(u,v),\qquad
&U=\sum_{i=1}^N \E_{i,N+1-i},\\
\RR_{21}(v,u;q)=\RR_{21}(u^{-1},v^{-1};q)=\RR_{12}(u,v;q^{-1}),\qquad &\RR_{12}(v^{-1},u^{-1})=\RR_{12}(u,v),
\end{split}
\end{equation}
where $^t$ denotes the transposition $(\E_{ij})^t= \E_{ji}$, $\forall i,j$. We have used auxiliary space notations, where the indices on an operator indicate in which space(s) it acts non trivially, and $\RR_{21}(u,v)=P_{12}\,\RR_{12}(u,v)\,P_{12}$ with $P_{12}$ the permutation of the two spaces 1 and 2.

\subsection{Iso- and anti- morphisms}
As one can easily deduce from the properties of the $\RR$-matrix \eqref{eq:symR}, there exist the
following morphisms from $U_q(\widehat{\mathfrak{gl}}_N)$ to $U_{q^{-1}}(\widehat{\mathfrak{gl}}_N)$.

\begin{prop}%\indent
\begin{itemize}
\item[(i)] The map $\vph$ defined by
\beq{phi}
\vph\big( T(u)\big)\ =\ U\,\tilde T^t(u)\,U^{-1},
\eeq
where $U$ is given in \eqref{eq:symR}, defines an isomorphism from $U_q(\widehat{\mathfrak{gl}}_N)$ to $U_{q^{-1}}(\widehat{\mathfrak{gl}}_N)$.
\item[(ii)] The map $\psi$ given by
\beq{psi}
\psi\big( T(u)\big)\ =\ \tilde T^t(u^{-1}),
\eeq
 defines an anti-isomorphism from $U_q(\widehat{\mathfrak{gl}}_N)$ to $U_{q^{-1}}(\widehat{\mathfrak{gl}}_N)$.
 In \r{phi} and \r{psi} $\LL(u)\in\Uq{N}$ and $\tilde\LL(u)\in\Uqin{N}$, respectively.
 \end{itemize}
\end{prop}
\proof
We start with the defining relation of $U_q(\widehat{\mathfrak{gl}}_N)$
\begin{equation}\label{rtt}
\RR_{12}(u,v;q)\cdot T_1(u) \cdot  T_2(v)= T_2(v)\cdot T_1(u)\cdot \RR_{12}(u,v;q),
\end{equation}
 apply the transpositions in space 1 and then in space 2, and conjugation by $U_1U_2$
 to get
\begin{equation}\label{rtt2}
U_1T_1^t(u)U_1^{-1}\cdot  U_2T_2^t(v)U_2^{-1}\cdot \RR_{12}(u,v;q)=
\RR_{12}(u,v;q)\cdot U_2T_2^t(v)U_2^{-1}\cdot U_1T_1^t(u)U_1^{-1},
\end{equation}
where we have used \eqref{eq:symR}. Using  {again \eqref{eq:symR}, we can rewrite \eqref{rtt2} as}
\begin{equation}\label{rtt3}
\vph\big(T_1(u)\big)\cdot \vph\big(T_2(v)\big)\cdot \RR_{21}(v,u;q^{-1})=
 \RR_{21}(v,u;q^{-1})\cdot \vph\big(T_2(v)\big)\cdot \vph\big(T_1(u)\big).
\end{equation}
After relabeling $1\leftrightarrow2$ and $u\leftrightarrow v$, one recognizes in \eqref{rtt3} the defining relations for
$U_{q^{-1}}(\widehat{\mathfrak{gl}}_N)$. This proves $(i)$.

Now, starting again from \eqref{rtt}, and applying transpositions in space 1 and then in space 2, and the transformation $(u,v)\to(u^{-1},v^{-1})$, we get
\begin{equation}\label{rtt2b}
T_1^t(u^{-1})\cdot  T_2^t(v^{-1})\cdot \RR_{21}(u^{-1},v^{-1})=
\RR_{21}(u^{-1},v^{-1})\cdot T_2^t(v^{-1})\cdot T_1^t(u^{-1}).
\end{equation}
Using once more \eqref{eq:symR}, it can be rewritten as
\begin{equation}\label{rtt3b}
\psi\big(T_1(u)\big)\cdot \psi\big(T_2(v)\big)\cdot \RR_{12}(u,v;q^{-1})=
 \RR_{12}(u,v;q^{-1})\cdot \psi\big(T_2(v)\big)\cdot \psi\big(T_1(u)\big).
\end{equation}
One recognizes in \eqref{rtt3b} the image of defining relations for
$U_{q^{-1}}(\widehat{\mathfrak{gl}}_N)$ under an anti-morphism, i.e. $\psi\big(T_1(u)\cdot T_2(v)\big)=\psi\big(T_2(v)\big)\cdot \psi\big(T_1(u)\big)$. 
This proves $(ii)$.
\finprf

\section{Closed formulas for dual off-shell BV}

Define the following combinations of the monodromy matrix elements
\begin{equation}\label{rrlbv11}
\begin{split}
\mathcal{C}^{\bar n}(\bar{t}_{\segg{\bar{n}}})&=\sum_{{\rm part}}
\prod_{k=1}^{N-1}\prod_{i,j\prec i',j'} \fun(\bar\t^k_{i',j'},\bar\t^k_{i,j})
\prod_{k=2}^{N-1}\Bigg(\prod_{i,j\prec i',j'} \fun(\bar \t^k_{i,j},\bar \t^{k-1}_{i',j'})
\prod_{i<j} \Izerr(\bar \t^k_{i,j}|\bar \t^{k-1}_{i,j})\Bigg)\\
&\times \prod_{k=2}^{N-1}\prod_{i,j\prec k,k}\LL_{k,k}(\bar\t^k_{i,j})
 \prod_{N-1\geq k\geq 1}^{\longleftarrow} \Bigg(\prod_{k<j\leq N}^{\longrightarrow}
\LL_{j,k}(\bar\t^k_{k,j-1})\Bigg)\,,
\end{split}
\end{equation}
\begin{equation}\label{rrlbv22}
\begin{split}
\wh{\mathcal{C}}^{\bar n}(\bar{t}_{\segg{\bar{n}}})&=\sum_{{\rm part}}
\prod_{k=1}^{N-1}\prod_{i,j\prect i',j'} \fun(\bar\t^k_{i',j'},\bar\t^k_{i,j})
\prod_{k=2}^{N-1}\Bigg(\prod_{i,j\prect i',j'} \fun(\bar \t^k_{i,j},\bar \t^{k-1}_{i',j'})
\prod_{i<j} \Izerl(\bar \t^k_{i,j}|\bar \t^{k-1}_{i,j})\Bigg)\\
&\times \prod_{k=1}^{N-2}\prod_{k,k\prect i,j}\LL_{k+1,k+1}(\bar\t^k_{i,j})
\prod_{1\leq k\leq N-1}^{\longrightarrow}\Bigg(\prod_{1\leq j\leq k}^{\longleftarrow}
\LL_{k+1,j}(\bar\t^k_{j,k})\Bigg)\,.
\end{split}
\end{equation}
We have the following
\begin{prop}\label{main-prop}\indent
\begin{itemize}
\item The morphism $\vph$ relates the universal off-shell pre-BV
\begin{equation}\label{phi-act}
\vph\Big(\wh{\mathcal{B}}_q^{\bar n}(\bar{t}_{\segg{\bar{n}}})\Big)=
{\mathcal{B}}_{q^{-1}}^{\,^\omega{\bar n}}(^\omega{\bar{t}_{\segg{\bar{n}}}})\,,
\end{equation}
where $\omega$ maps the  sets of Bethe parameters into the fully permuted sets:
\begin{equation*}
\omega:
{\bar{t}_{\segg{\bar{n}}}}
\to\ ^\omega{\bar{t}_{\segg{\bar{n}}}}=\left\{
t^{N-1}_{1},\ldots,t^{N-1}_{n_{N-1}};
t^{N-2}_{1},\ldots, t^{N-2}_{n_{N-2}};
\ldots\ldots;
t^{2}_{1},\ldots,t^{2}_{n_{2}};
t^{1}_{1},\ldots,t^{1}_{n_{1}}
\right\}\,,
\end{equation*}
and accordingly for the sets of cardinalities:
\begin{equation*}
\omega: \bar n\to\ ^\omega{\bar n}=\{n_{N-1},n_{N-2},\ldots,n_2,n_1\}\,.
\end{equation*}
\item The combinations \r{rrlbv11} and \r{rrlbv22} are related to the pre-BV
\r{bvp1} and \r{bvp2} by the antimorphism $\psi$
\begin{equation}\label{psi-act}
\psi\Big(\mathcal{B}_q^{\bar n}(\bar{t}_{\segg{\bar{n}}})\Big)=
\mathcal{C}_{q^{-1}}^{\bar n}(\bar{t}^{-1}_{\segg{\bar{n}}})
\mb{and}
\psi\Big(\wh{\mathcal{B}}_q^{\bar n}(\bar{t}_{\segg{\bar{n}}})\Big)=
\wh{\mathcal{C}}_{q^{-1}}^{\bar n}(\bar{t}^{-1}_{\segg{\bar{n}}})\,,
\end{equation}
where
\begin{equation*}
{\bar{t}^{-1}_{\segg{\bar{n}}}}=\left\{
\sk{t^{1}_{1}}^{-1},\ldots,\sk{t^{1}_{n_{1}}}^{-1};
\sk{t^{2}_{1}}^{-1},\ldots,\sk{t^{2}_{n_{2}}}^{-1};
\ldots\ldots;
\sk{t^{N-1}_{1}}^{-1},\ldots,\sk{t^{N-1}_{n_{N-1}}}^{-1}
\right\}
\end{equation*}
is the set of the inverses of Bethe parameters.
\end{itemize}
\end{prop}
\proof
 We start with the first item.  We write expression for pre-BV \r{bvp1}  for
 the algebra $U_{q^{-1}}(\widehat{\mathfrak{gl}}_N)$ and using the sets
$\bar n\leftrightarrow\ ^\omega{\bar n}$,
${\bar{t}_{\segg{\bar{n}}}}\leftrightarrow\ ^\omega{\bar{t}_{\segg{\bar{n}}}}$:
\begin{equation}\label{BVP1}
\begin{split}
&\mathcal{B}^{\,^\omega\bar n}_{q^{-1}}(^\omega\bar{t}_{\segg{\bar{n}}})=\sum_{{\rm part}}
\prod_{k=1}^{N-1}\prod_{i,j\prec i',j'} \fun_{q^{-1}}(\bar\t^{N-k}_{i',j'},\bar\t^{N-k}_{i,j})
\prod_{k=2}^{N-1}\prod_{i,j\prec i',j'} \fun_{q^{-1}}(\bar \t^{N-k}_{i,j},\bar \t^{N-k+1}_{i',j'})\\
&\times \prod_{k=2}^{N-1}\prod_{i<j} \Izerl_{q^{-1}}(\bar \t^{N-k}_{i,j}|\bar \t^{N-k+1}_{i,j})
 \prod_{1\leq k\leq N-1}^{\longrightarrow} \Bigg(\prod_{N\geq j> k}^{\longleftarrow}
\tilde\LL_{k,j}(\bar\t^{N-k}_{k,j-1})\Bigg)\prod_{k=2}^{N-1}\prod_{i,j\prec k,k}\tilde\LL_{k,k}(\bar\t^{N-k}_{i,j}).
\end{split}
\end{equation}
Here, $\tilde\LL_{i,j}(t)$ are the matrix elements of the $\Uqin{N}$ monodromy matrix.

The next step is to use the formulas \r{prop-fct} to change the rational functions $\fun_{q^{-1}}\to\fun_q$
and $\Izerl_{q^{-1}}\to \Izerr_q$, and apply the morphism
$\vph^{-1}$ to obtain, starting  from
 \r{BVP1}:
\begin{equation}\label{BVP2}
\begin{split}
\vph^{-1}\sk{\mathcal{B}^{\,^\omega\bar n}_{q^{-1}}(^\omega\bar{t}_{\segg{\bar{n}}})}=\sum_{{\rm part}}&
\prod_{k=1}^{N-1}\prod_{i,j\prec i',j'} \fun_{q}(\bar\t^{N-k}_{i,j},\bar\t^{N-k}_{i',j'})\\
&\times \prod_{k=2}^{N-1}\prod_{i,j\prec i',j'} \fun_{q}(\bar \t^{N-k+1}_{i',j'},\bar \t^{N-k}_{i,j})\\
&\times \prod_{k=2}^{N-1}\prod_{i<j} \Izerr_{q}(\bar \t^{N-k+1}_{i,j}|\bar \t^{N-k}_{i,j})\\
&\times \prod_{1\leq k\leq N-1}^{\longrightarrow} \Bigg(\prod_{N\geq j> k}^{\longleftarrow}
\LL_{N+1-j,N+1-k}(\bar\t^{N-k}_{k,j-1})\Bigg)\\
&\times\prod_{k=2}^{N-1}\prod_{i,j\prec k,k}\LL_{N+1-k,N+1-k}(\bar\t^{N-k}_{i,j})\,.
\end{split}
\end{equation}
Let us consider the products in each line of \r{BVP2} separately.
\begin{itemize}
\item In the product of the first line we renumber the partition \r{pt11} of the set
$\bar{t}_{\segg{\bar{n}}}$. We introduce new pairs of  integers $\hj=N-i$ and $\hi=N-j$ which
can be also used to numerate the partition \r{pt11} since it is clear that for any given $k$ which
defines the pair $i,j$ the new integers $\hi,\hj$ will also satisfy the condition
$1\leq\hi\leq N-k\leq\hj\leq N-1$. One can verify that the condition  $i,j\prec i',j'$ will be
reformulated as $\hi',\hj'\prect \hi,\hj$. Then, after renaming the integers:
$k\to N-k$, $\hi'\to i$, $\hj'\to j$, $\hi\to i'$ and $\hj\to j'$, one can see that the first product in \r{BVP2}
coincides literally with the first product in \r{bvp2}
\begin{equation*}
\prod_{k=1}^{N-1}\prod_{i,j\prect i',j'} \fun(\bar\t^k_{i',j'},\bar\t^k_{i,j})\,.
\end{equation*}
\item Using the same arguments, the same change of dumb integers and
renaming $k\to N-k+1$, $\hi'\to i$, $\hj'\to j$, $\hi\to i'$ and $\hj\to j'$,
we can prove that the second product in \r{BVP2} is equal to
\begin{equation*}
\prod_{k=2}^{N-1}\prod_{i,j\prect i',j'} \fun(\bar \t^k_{i,j},\bar \t^{k-1}_{i',j'}).
\end{equation*}
\item In the third product of \r{BVP2} we change numeration of the partition of the set
$\bar t^{N-k+1}$ in such a way to obtain the product
\begin{equation*}
\prod_{k=2}^{N-1}
\prod_{i<j} \Izerr(\bar \t^k_{i,j}|\bar \t^{k-1}_{i,j}).
\end{equation*}
 \item Renaming the partitions $\{\bar t^{N-k}_{k,j-1}\}\to\{\bar t^{N-k}_{N-j+1,N-k}\}$ and
 changing the dumb integers $k\to N-k$ and $j\to N-j+1$, we obtain for the forth product in \r{BVP2}
 \begin{equation*}
 \prod_{N-1\geq k\geq 1}^{\longleftarrow}\Bigg(\prod_{1\leq j\leq k}^{\longrightarrow}
\LL_{j,k+1}(\bar\t^k_{j,k})\Bigg).
 \end{equation*}
 \item Finally in the fifth product of \r{BVP2}, we renumber the partition
 $\{\bar t^{N-k}_{i,j}\}\to\{\bar t^{N-k}_{N-j,N-i}\}$ and rename the dumb
 variables $k\to N-k$, $N-j\to i$ and $N-i\to j$. It  leads to the product
 \begin{equation*}
\prod_{k=1}^{N-2}\prod_{k,k\prect i,j}\LL_{k+1,k+1}(\bar\t^k_{i,j})\,.
 \end{equation*}
\end{itemize}
This proves the relation \r{phi-act} of the first item of the Proposition.

For the proof of the second item we consider the first relation in \r{psi-act}. The second one can be proved analogously.
We rewrite again the expression of the pre-BV \r{bvp1}  for
 $U_{q^{-1}}(\widehat{\mathfrak{gl}}_N)$ and
${\bar{t}_{\segg{\bar{n}}}}\leftrightarrow\ \sk{\bar{t}_{\segg{\bar{n}}}}^{-1}$
\begin{equation}\label{BVP3}
\begin{split}
&\mathcal{B}^{\bar n}_{q^{-1}}\sk{\sk{\bar{t}_{\segg{\bar{n}}}}^{-1}}=\sum_{{\rm part}}
\prod_{k=1}^{N-1}\prod_{i,j\prec i',j'} \fun_{q^{-1}}(({\bar\t^{k}_{i',j'}})^{-1},(\bar\t^{k}_{i,j})^{-1})
\prod_{k=2}^{N-1}\prod_{i,j\prec i',j'} \fun_{q^{-1}}((\bar \t^{k}_{i,j})^{-1},(\bar \t^{k-1}_{i',j'})^{-1})\\
&\times \prod_{k=2}^{N-1}\prod_{i<j} \Izerl_{q^{-1}}((\bar \t^{k}_{i,j})^{-1}|(\bar \t^{k-1}_{i,j})^{-1})
 \prod_{1\leq k\leq N-1}^{\longrightarrow} \Bigg(\prod_{N\geq j> k}^{\longleftarrow}
\tilde\LL_{k,j}((\bar\t^{k}_{k,j-1})^{-1})\Bigg)\prod_{k=2}^{N-1}\prod_{i,j\prec k,k}\tilde\LL_{k,k}((\bar\t^{k}_{i,j})^{-1})
\,.
\end{split}
\end{equation}
Here again $\tilde\LL_{i,j}(t)$ are the matrix elements of the $\Uqin{N}$ monodromy matrix.
Now we apply the antimorphism $\psi$ to \r{BVP3}, using once more the relation \r{prop-fct} to obtain
\begin{equation}\label{BVP4}
\begin{split}
&\psi\sk{\mathcal{B}^{\bar n}_{q^{-1}}\sk{\sk{\bar{t}_{\segg{\bar{n}}}}^{-1}}}=\sum_{{\rm part}}
\prod_{k=1}^{N-1}\prod_{i,j\prec i',j'} \fun_{q}({\bar\t^{k}_{i',j'}},\bar\t^{k}_{i,j})
\prod_{k=2}^{N-1}\prod_{i,j\prec i',j'} \fun_{q}(\bar \t^{k}_{i,j}),\bar \t^{k-1}_{i',j'})\\
&\times \prod_{k=2}^{N-1}\prod_{i<j} \Izerr_{q}(\bar \t^{k}_{i,j}|\bar \t^{k-1}_{i,j})
\prod_{k=2}^{N-1}\prod_{i,j\prec k,k}\LL_{k,k}(\bar\t^{k}_{i,j})
\prod_{N-1\geq k\geq 1}^{\longleftarrow} \Bigg(\prod_{k<j\leq N}^{\longrightarrow}
\LL_{j,k}(\bar\t^{k}_{k,j-1})\Bigg)\,.
\end{split}
\end{equation}
In \r{BVP4} the property of the antimorphism $\psi$ is used to reverse the order in the product of the non-commuting operators.
The r.h.s. of \r{BVP4} coincides with the combination \r{rrlbv11}. Thus, the second item of the Proposition is proved.
\finprf

Repeating arguments used for the proof of the Proposition~\ref{main-prop}, we can deduce from properties of
$\bbb^{\bar n}(\bar{t}_{\segg{\bar{n}}})$, similar properties for
$\wh\bbb_q^{\bar n}(\bar{t}_{\segg{\bar{n}}})$ and $\ccc_q^{\bar n}(\bar{t}_{\segg{\bar{n}}})$. As an illustration, we prove the following.

\begin{prop}\label{LBV-BE}
The combinations of the monodromy matrix elements
\r{rrlbv11} and \r{rrlbv22} are the dual pre-BV because their left action onto vacuum vector
defined by \r{lvec} produces the dual off-shell BV
\begin{equation}\label{duals}
\ccc^{\bar n}(\bar{t}_{\segg{\bar{n}}})=\lvec \mathcal{C}^{\bar n}(\bar{t}_{\segg{\bar{n}}})=
\lvec \wh{\mathcal{C}}^{\bar n}(\bar{t}_{\segg{\bar{n}}})\,,
\end{equation}
which are eigenvectors under
the left action of the transfer matrix $\Ht(t)=\sum_{i=1}^N\LL_{i,i}(t)$:
\begin{equation}\label{dBVprop}
\ccc^{\bar n}(\bar{t}_{\segg{\bar{n}}})\tau(t;\bar{t}_{\segg{\bar{n}}})
=\ccc^{\bar n}(\bar{t}_{\segg{\bar{n}}})\Ht(t)\,,
\end{equation}
with eigenvalue
\begin{equation}\label{eigenv}
\tau(t;\bar{t}_{\segg{\bar{n}}})
=\sum_{i=1}^N \lambda^+_i(t)\prod_{j=1}^{n_{i-1}}\fun(t,t^{i-1}_j)
\prod_{j=1}^{n_{i}}\fun(t^i_j,t)\,,
\end{equation}
provided the Bethe equations
\begin{equation}\label{BEq}
\frac{\lambda_i(t^i_j)}{\lambda_{i+1}(t^i_j)}=(-1)^{n_i-1}
\prod_{m=1\atop m\neq j}^{n_i}
\frac{\fun(t^i_j,t^{i}_m)}{\fun(t^{i}_m,t^i_j)}\
\prod_{m=1}^{n_{i-1}}
\fun(t^i_j,t^{i-1}_m)^{-1}
\prod_{m=1}^{n_{i+1}}
\fun(t^{i+1}_m,t^i_j)
\end{equation}
are satisfied.
\end{prop}
\proof
We start with the relation \r{BV1}, written in the algebra $\Uqin{N}$ and for the set of parameters
${\bar{t}^{-1}_{\segg{\bar{n}}}}$. Then, we apply $\psi^{-1}$ to get \r{duals}.

For the remaining part of the proposition, we use  a result proved in the paper \cite{FKPR}. It states that
in $\Uq{N}$, we have
\begin{equation}\label{reig-vec}
\Ht(t)\bbb^{\bar n}(\bar{t}_{\segg{\bar{n}}})=\tau(t;\bar{t}_{\segg{\bar{n}}})
\bbb^{\bar n}(\bar{t}_{\segg{\bar{n}}})\,,
\end{equation}
if the Bethe equations \r{BEq} are satisfied. We reformulate this result for $\Uqin{N}$,  in the following form:
\begin{eqnarray}\label{reig-vec2}
\Ht_{q^{-1}}(t)\bbb^{\bar n}_{q^{-1}}({\bar{t}_{\segg{\bar{n}}}})&=&\tau_{q^{-1}}(t;{\bar{t}_{\segg{\bar{n}}}})
\bbb_{q^{-1}}^{\bar n}({\bar{t}_{\segg{\bar{n}}}})
\nonu
&+&
\sum_{i=1}^{N-1}\sum_{j=1}^{n_i} \cO_{ij}\Big\{ \prod_{m=1\atop m\neq j}^{n_i}
\fun_{q^{-1}}(t^{i}_m,t^i_j)
\prod_{m=1}^{n_{i-1}}
\fun_{q^{-1}}(t^i_j,t^{i-1}_m)
\ \tilde T_{i,i}(t^i_j)
\nonu
&-& (-1)^{n_i-1}
\prod_{m=1\atop m\neq j}^{n_i}
{\fun_{q^{-1}}(t^i_j,t^{i}_m)}{}\
\prod_{m=1}^{n_{i+1}}
\fun_{q^{-1}}(t^{i+1}_m,t^i_j)\ \tilde T_{i+1,i+1}(t^i_j)
\Big\}\rvec\,,
\end{eqnarray}
where $\Ht_{q^{-1}}(t)=\sum_{i=1}^{N} \tilde\LL_{ii}(t)$ is the transfer matrix in $\Uqin{N}$, and $\cO_{ij}$ are some operators, whose explicit form is not needed for our proof.

A direct calculation, using properties \r{prop-fct}, shows that $\tau_{q^{-1}}(t^{-1};{\bar{t}^{-1}_{\segg{\bar{n}}}})=\tau_q(t;\bar{t}_{\segg{\bar{n}}})$.
It is also clear that $\psi^{-1}(\Ht_{q^{-1}}(t^{-1}))=\Ht_q(t)$. Then, applying $\psi^{-1}$ to
\r{reig-vec2} written using a spectral parameter $t^{-1}$ and the sets
${\bar{t}^{-1}_{\segg{\bar{n}}}}$, we get in $\Uq{N}$
\begin{eqnarray}
\ccc^{\bar n}(\bar{t}_{\segg{\bar{n}}})\Ht(t)
&=&\ccc^{\bar n}(\bar{t}_{\segg{\bar{n}}})\tau(t;\bar{t}_{\segg{\bar{n}}})
\nonu
&+&\sum_{i=1}^{N-1}\sum_{j=1}^{n_i}
\lvec \Big\{\prod_{m=1\atop m\neq j}^{n_i}
\fun_{q}(t^{i}_m,t^i_j)
\prod_{m=1}^{n_{i-1}}
\fun_{q}(t^i_j,t^{i-1}_m)
\  T_{i,i}(t^i_j)
\nonu
&&
- (-1)^{n_i-1}
\prod_{m=1\atop m\neq j}^{n_i}
\fun_{q}(t^i_j,t^{i}_m)\
\prod_{m=1}^{n_{i+1}}
\fun_{q}(t^{i+1}_m,t^i_j)\  T_{i+1,i+1}(t^i_j)
 \Big\}\ \psi^{-1}(\cO_{ij})\,,
\end{eqnarray}
where we have used once more the relations \r{prop-fct}. This {ends} the proofs.
 \finprf

\subsection{BV for $\Uq{3}$-symmetric integrable models}

In the paper \cite{BPRS-trigo} the action of the monodromy matrix elements onto right off-shell Bethe
vectors was calculated in the framework of the current approach and the formulas for the dual off-shell
BV were announced. Let us verify that the formulas used in \cite{BPRS-trigo} can be obtained
from the generic formulas proved in the present paper.

In order to rewrite the $\Uq{3}$ off-shell BV in the form used in our previous papers we rename
the sets of Bethe parameters
\begin{equation*}
\bar t^1,\bar t^2 \to \bar u,\bar v\,.
\end{equation*}
We rename as follows the subsets of these sets
\begin{equation}\label{rename}
\begin{split}
\bar t^1=\bar t^1_{1,1}\cup\bar t^1_{1,2}\to \bar u_{\st}\cup\bar u_{\so}=\bar u\,,\\
\bar t^2=\bar t^2_{1,2}\cup\bar t^2_{2,2}\to \bar v_{\so}\cup\bar v_{\st}=\bar v\,.
\end{split}
\end{equation}
With this renaming of the parameters and the partitions formulas \r{bvp1} and \r{bvp2} yield
the following expressions for the off-shell BV:
\begin{equation}\label{rBV3}
\begin{split}
\bbb(\bar u,\bar v)&=\sum_{\rm part}
\Izerl(\bar v_{\so}|\bar u_{\so}) \fun(\bar u_{\so},\bar u_{\st})\fun(\bar v_{\st},\bar v_{\so})
\LL_{13}(\bar u_{\so})\LL_{12}(\bar u_{\st})\LL_{23}(\bar v_{\st})\ll_2(\bar v_{\so})\rvec\,,\\
\bbb(\bar u,\bar v)&=\sum_{\rm part}
\Izerr(\bar v_{\so}|\bar u_{\so}) \fun(\bar u_{\so},\bar u_{\st})\fun(\bar v_{\st},\bar v_{\so})
\LL_{13}(\bar v_{\so})\LL_{23}(\bar v_{\st})\LL_{12}(\bar u_{\st})\ll_2(\bar u_{\so})\rvec\,,
\end{split}
\end{equation}
respectively. The second formula in \r{rBV3} coincides literally with the formula (5.1) in \cite{BPRS-trigo}
after changing overall normalization of the BV
\begin{equation}\label{ch-nor}
\bbb(\bar u,\bar v)\to \frac{\bbb(\bar u,\bar v)}{\fun(\bar v,\bar u)\ll_2(\bar u)\ll_2(\bar v)}\,.
\end{equation}
This change of normalization of the BV is useful for the calculation of the action of the
monodromy matrix elements onto them because it remove certain poles in the BV and makes
formulas for this action quite effective (see details in \cite{BPRS-trigo}).
The first formula in \r{rBV3} yields an alternative expression for the off-shell BV and their equivalence
can be proved directly from the $RTT$ commutation relations \r{YB} and certain identities for the Izergin determinants.

Analogously, formulas \r{rrlbv11} and \r{rrlbv22} yield the following expressions for the dual or left
off-shell BV
\begin{equation}\label{lBV3}
\begin{split}
\ccc(\bar u,\bar v)&=\sum_{\rm part}
\Izerr(\bar v_{\so}|\bar u_{\so}) \fun(\bar u_{\so},\bar u_{\st})\fun(\bar v_{\st},\bar v_{\so})
\ll_2(\bar v_{\so})\lvec
\LL_{32}(\bar v_{\st})\LL_{21}(\bar u_{\st})\LL_{31}(\bar u_{\so})\,,\\
\ccc(\bar u,\bar v)&=\sum_{\rm part}
\Izerl(\bar v_{\so}|\bar u_{\so}) \fun(\bar u_{\so},\bar u_{\st})\fun(\bar v_{\st},\bar v_{\so})
\ll_2(\bar u_{\so})\lvec
\LL_{21}(\bar u_{\st})\LL_{32}(\bar v_{\st})\LL_{31}(\bar v_{\so})\,,
\end{split}
\end{equation}
and after the same change of the normalization as in \r{ch-nor}, the second formula in \r{lBV3}
coincides literally with formula (5.2) in the paper \cite{BPRS-trigo}.

\section*{Conclusion}

In this paper we have obtained explicit formulas for the right and left (dual) off-shell BV
 in the form of  sums over partitions of the sets of
Bethe parameters using morphisms of the algebra $\Uq{N}$.
Our starting formulas were presentations of the off-shell BV
in terms of sums over permutations in the sets of  Bethe parameters obtained previously
in \cite{KhP-Kyoto,OPS} by the current approach. Formulas for left or dual off-shell Bethe
vectors are necessary to address the problem of calculation of the scalar products of the
BV.

In a previous paper \cite{BPRS-trigo}, we computed the action of the monodromy matrix
elements onto nested off-shell BV in integrable models with $GL(3)$ trigonometric $R$-matrix.
The morphisms \r{phi} and antimorphism \r{psi} introduced in this paper allows one to relate easily
different formulas of these actions. For example, one can obtain using antimorphism $\psi$
the left action of the monodromy matrix
elements onto dual off-shell BV from the corresponding  formulas of the right action
onto right off-shell BV and many other useful relations.

\section*{Acknowledgements}
We warmly thank S. Belliard for his contribution at the early stage of this work.
Work of S.P. was supported in part by RFBR grant 11-01-00980-a and  grant
of Scientific Foundation of NRU HSE 12-09-0064. E.R. was supported by ANR Project
DIADEMS (Programme Blanc ANR SIMI1 2010-BLAN-0120-02).
N.A.S. was  supported by the Program of RAS Basic Problems of the Nonlinear Dynamics,
RFBR-11-01-00440-a, RFBR-13-01-12405-ofi-m2, SS-4612.2012.1.

\end{document}